\documentclass[11pt]{article}
\usepackage{jheppub}
\usepackage{amsmath,amssymb,amsfonts,amsbsy}

\usepackage{hyperref}

\newcommand{\nc}{\newcommand}
\nc{\rnc}{\renewcommand }
\nc{\tep}{\tilde{\epsilon}}
\rnc{\d}{\mathrm{d}}
\nc{\D}{\partial}
\nc{\bg}{\bar{g}}
\nc{\g}{\gamma}
\rnc{\o}{\omega}
\nc{\n}{{(n)}}
\rnc{\t}{\tau}
\nc{\ep}{\epsilon}
\nc{\K}{\mathcal{K}}
\nc{\cK}{\mathcal{K}}
\nc{\J}{\mathcal{J}}
\rnc{\H}{{\mathcal{H}}}
\nc{\lrarrow}{\leftrightarrow}
\nc{\nn}{\nonumber}
\nc{\tto}{\rightarrow}
\nc{\be}{\begin{equation}}
\nc{\ee}{\end{equation}}
\nc{\bea}{\begin{eqnarray}}
\nc{\eea}{\end{eqnarray}}
\rnc{\(}{\left(}
\rnc{\)}{\right)}
\nc{\half}{\frac{1}{2}}

\nc{\pa}{\partial}
\nc{\s}{\sigma}
\nc{\<}{\langle}
\rnc{\>}{\rangle}

\def\td{\tilde}

\title{\mbox{}\\ \vspace{1cm} Holographic realization of gauge mediated supersymmetry breaking}

\author[a,b]{Kostas Skenderis}
\author[b]{and Marika Taylor}

\affiliation[a]{KdV Institute for Mathematics, }
\affiliation[b]{Institute for Theoretical Physics, \\
Science Park 904, 1090 GL Amsterdam, the Netherlands.}

\emailAdd{K.Skenderis@uva.nl}
\emailAdd{M.Taylor@uva.nl}

\abstract{
The general gauge mediation scenario provides a framework in which properties of a visible sector
with soft supersymmetry breaking
are computed from current correlation functions in the supersymmetry breaking hidden sector. In this paper we will use
holography to model strongly coupled hidden sectors by weakly curved geometries and describe how the current correlators
relevant for general gauge mediation are computed by holographic methods.
We illustrate the general setup by a toy example which captures most of the relevant features.
}

\begin{document}
\maketitle


\section{Introduction}

Low energy supersymmetry is a prime candidate for physics beyond the Standard Model as it addresses
the hierarchy problem and predicts gauge coupling unification. The minimal supersymmetric standard model
(MSSM) with soft supersymmetry breaking terms has a large number of parameters, which are already
tightly constrained by experimental limits on flavor changing neutral currents and CP violation. If supersymmetry
is discovered at LHC, the challenge will be to construct theoretical models of supersymmetry breaking and its
mediation to the visible sector which reproduce the observed soft parameters.

Gauge mediation is a popular method of supersymmetry breaking in which the susy breaking is communicated to the
visible sector via gauge interactions, see \cite{Giudice:1998bp} for a review. Gauge mediated supersymmetry breaking has both advantages
such as automatic flavor universality (cf gravity mediation) and disadvantages such as the absence of a satisfactory and
natural mechanism to generate $\mu$ and $B_{\mu}$ parameters of the right scale to generate electroweak
symmetry breaking. Clearly it would be desirable to develop sharp criteria for the
features a phenomenologically acceptable gauge mediated model should have. Recently progress was made in this direction in
\cite{Meade:2008wd} where direct gauge mediation was given a general, model-independent definition. The basic idea is to
consider models in which the theory decouples into the MSSM and a separate susy breaking hidden sector as the MSSM
gauge couplings go to zero; in such scenarios the visible sector gauge group must be part of a weakly gauged global symmetry
$G$ of the hidden sector. Then, as will be reviewed below,
the leading order dependence of the soft masses follows from the properties of the current-current correlators of $G$. This
framework was denoted ``General Gauge Mediation" (GGM) in \cite{Meade:2008wd}
and was further developed in, for example, \cite{Buican:2008ws,Komargodski:2008ax}.

The key advantage of the GGM framework is that it makes manifest properties which are generic to gauge mediated
scenarios whilst providing sharp definitions of non-generic properties.
Whilst the GGM framework assumes that the visible and hidden sectors decouple as the (MSSM) gauge couplings tend to zero,
it does not assume that the hidden sector is itself weakly coupled: the formulation of gauge mediation
in terms of current correlators applies equally well to strongly coupled hidden sectors. A natural question is thus whether
one can find phenomenologically interesting models in which the hidden sector is strongly coupled. In this paper we will show
how gravity/gauge theory duality \cite{Maldacena:1997re, Gubser:1998bc,Witten:1998qj} can be used to model a strongly
coupled hidden sector by a weakly curved geometry.
We will discuss what features the geometry should have in order to be dual
to a theory with spontaneously broken supersymmetry, and we will show how holographic methods can be used to compute the
current correlators relevant for GGM. We illustrate the general setup by a specific toy example which
neatly demonstrates the holographic approach. Earlier work on holographic gauge mediation can be found in
\cite{Benini:2009ff,McGuirk:2009am,McGuirk:2011yg}.

\section{General gauge mediation}

In the basic setup of general gauge mediation \cite{Meade:2008wd}, one considers a renormalizable hidden sector where the
supersymmetry is broken spontaneously. The hidden sector should have a global symmetry group $G$ which
contains the standard model (SM) gauge symmetry and the only coupling to the visible sector should occur via
SM gauge interactions (so that the sectors decouple as $g_{SM} \rightarrow 0$). In this setup information
about the visible sector is encoded in correlation functions of the current superfield of the group $G$;
in particular, soft masses in the visible sector are encoded in two point functions. Here we will review
the relevant features of generalized gauge mediation, in particular, identifying the quantities that one
would like to compute.

For simplicity let us first discuss the case where $G = U(1)$ as the generalization to nonabelian groups
is straightforward. Working in ${\cal N} = 1$ superspace, the current superfield ${\cal J}$ is a real
linear superfield defined by
\be
D^2 {\cal J} = \bar{D}^2 {\cal J} = 0,
\ee
and in components it takes the form:
\be
{\cal J} = J + i \theta j - \i \bar{\theta} \bar{j} - \theta \sigma^{\mu} \bar{\theta}
j_{\mu} + \half \theta^2 \bar{\theta} \bar{\sigma}^{\mu} \pa_{\mu} j + \cdots \label{comp}
\ee
with $\pa^{\mu} j_{\mu} = 0$. Note that
in the case of $U(1)$ the one point function $\< J \>$ can in principle be
non-zero, $\< J \> = \zeta$, although such an effective FI term is undesirable as it can lead to
non-positive definite contributions to visible sector masses. In what follows we will assume such a term
is absent. The nonzero current two point functions in momentum space are:
\bea
\< J(k) J(-k) \> &=& \td{C}_{0}(k^2/M^2); \label{con} \\
\< j_{\alpha} (k) \bar{j}_{\dot{\alpha}} (-k) \> &=& \sigma^{\mu}_{\alpha \dot{\alpha}}
k_{\mu} \td{C}_{1/2} (k^2/M^2); \nn \\
\< j_{\mu} (k) j_{\nu} (-k) \> &=&
- k^2 \pi_{\mu \nu} \td{C}_1 (k^2/M^2); \nn \\
\< j_{\alpha} (k) j_{\beta} (-k) \> &=& \ep_{\alpha \beta} M \td{B}_{1/2} (k^2/M^2), \nn
\eea
where $\pi_{\mu \nu} = \eta_{\mu \nu}  - k_{\mu} k_{\nu}/k^2$ and $M$ is a characteristic scale of the theory.
If supersymmetry is unbroken then \cite{Meade:2008wd}
\be \label{susy-relation}
\td{C}_{0} = \td{C}_{1/2} = \td{C}_{1}; \qquad \td{B}_{1/2} = 0.
\ee
When supersymmetry is spontaneously broken these relations should hold
at leading order in the short distance ($k \to \infty$) limit.

The current superfield of the hidden sector is then coupled to the vector superfield ${\cal V}$
of the visible sector via the coupling:
\bea
L_{\int} &=& 2g \int d^4 \theta {\cal J} {\cal V} + \cdots \\
&=& g (J D - \lambda j - \bar{\lambda} \bar{j} - j^{\mu} V_{\mu}) + \cdots \nn
\eea
where we work in Wess-Zumino gauge and ellipses denote terms of order $g^2$ required by gauge invariance.
In the latter expression the vector superfield is expanded in its components, the gauge field $V_{\mu}$,
the gaugino $\lambda_{\alpha}$ etc. Such a coupling is required for all gauge fields in the visible sector,
namely the complete $U(1) \times SU(2) \times SU(3)$ or GUT symmetry group, and thus
${\cal J} \rightarrow {\cal J}^{(r)}$, with $r$ denoting the visible sector gauge group.

When one couples the hidden sector to the visible sector by weakly gauging the global symmetry in this way the
soft masses generated at leading order in the gauge coupling $g$ (after integrating out the hidden sector)
are given by:
\be
m_{\lambda^{(r)}} = g^2 M \td{B}^{(r)}_{1/2} (0)
\ee
for the gauginos. From (\ref{susy-relation}) one sees that the gaugino mass is generated at tree level whenever the
hidden sector breaks supersymmetry.
For the sfermions the masses are generated at one loop and are given by
\bea
m^2_{\td{f}} &=& g^4 \sum_{r} c (\td{f}, r) M^2 {\cal A}^{(r)}; \\
{\cal A}^{(r)} &=& \int \frac{d^4 q}{(2 \pi)^4} \frac{1}{q^2} \left (3 \td{C}^{(r)}_1(q) - 4 \td{C}^{(r)}_{1/2}(q)
+ \td{C}^{(r)}_0 (q) \right ), \nn
\eea
where $c(\td{f},r)$ denotes the quadratic Casimir of the sfermion representation
of the gauge group. The masses vanish when the hidden sector is supersymmetric, see (\ref{susy-relation}),
and the soft susy breaking masses for given generations are related by sum rules.
Note that, whilst the gaugino mass is determined by the zero momentum limit of
the current, computation of the sfermion masses requires knowledge of the full momentum dependence of
the currents.

\section{Holographic realization}

As mentioned previously the general gauge mediation framework
does not assume that the hidden sector is itself weakly coupled. Gravity/gauge theory
duality has the potential to provide a large class of examples of strongly coupled hidden sectors:
the idea would be to find weakly curved geometries which are holographically dual to theories with
spontaneously broken supersymmetry.

Basic features required of a holographic modelling of a hidden sector are:
\begin{enumerate}
\item{The geometry is dual to a supersymmetry breaking state of a supersymmetric field theory.}
\item{The supersymmetry breaking state should be at least metastable, which corresponds in the geometry to perturbative
stability.}
\item{The supersymmetry field theory should admit global symmetry, with the two point functions of the global currents
being computable by  studying fluctuations of gauge fields in the holographically dual geometry.}
\end{enumerate}
A priori one need not restrict the UV behavior of the supersymmetric field theory; in what follows our toy example
will be UV conformal, but precision holography has been developed well beyond asymptotically AdS spacetimes (and hence
UV conformal theories).

Finding holographic duals to (metastable) supersymmetric breaking states may at first sight seem challenging, but in fact
{\it fake supersymmetry} provides a new technique to construct such backgrounds.
The main claim is that appropriate holographic backgrounds can be obtained by considering $(d+1)$-dimensional
non-supersymmetric domain wall solutions to gravity/scalar theories, suitably embedded into $(d+1)$ or ten dimensional
supergravity.

Let us consider the properties of holographic backgrounds describing (meta)stable ground states of the dual
$d$-dimensional field theories in which supersymmetry is spontaneously broken. The corresponding holographic duals
should involve $(d+1)$ non-compact dimensions and, since
the ground states of interest preserve Poincar\'{e} invariance in $d$ dimensions,
the effective $(d+1)$-dimensional solutions should be domain wall solutions such that:
\be
ds^2 = e^{2 A(r)} \eta_{\mu \nu} dx^{\mu} dx^{\nu} + dr^2; \qquad \phi^a = \phi^a(r),
\ee
where $\eta_{\mu \nu}$ is the $d$-dimensional Minkowski metric and the scalars $\phi^a$ depend only on
the radial coordinate. Note that when $A=r$ with the scalars constant the solution is anti-de Sitter space,
and thus the dual theory is conformal. Such domain walls arise as solutions of an action involving
gravity coupled to scalar fields in $(d+1)$ dimensions:
\be
S = \frac{1}{2 \kappa^2 } \int d^{d+1} x \sqrt{-g} \left [ R - \frac{1}{2} g_{ab} (\phi^c) \pa_{m} \phi^a
\pa^{m} \phi^b - V(\phi^a) \right ]. \label{g-sc}
\ee
Actions of this type arise from truncations of gauged supergravity theories in $(d+1)$ dimensions, which in
turn can be interpreted as consistent Kaluza-Klein reductions of 10 dimensional string backgrounds.
The most relevant examples here are dualities between $AdS_5 \times X^5$ where $X^5$ is Sasaki-Einstein and
${\cal N} = 2$ SCFTs. Kaluza-Klein reduction of the ten-dimensional supergravity theory on $X^5$ results in
a five-dimensional theory, which is believed to admit a consistent truncation to an ${\cal N} =2$
gauged supergravity.

Let us restrict to the case of a single active scalar $\phi$, the generalization to additional scalars being
straightforward; then the Einstein and scalar equations of motion become:
\bea
\dot{A}^2 &=& \frac{1}{2 d(d-1)} \dot{\phi}^2 - \frac{1}{d(d-1)} V(\phi); \label{feq} \\
\ddot{A} &=& - \frac{1}{2 (d-1)} (\dot{\phi})^2; \qquad
\ddot{\phi} + d \dot{A} \dot{\phi} = \frac{\partial V}{\partial \phi}, \nn
\eea
where $\dot{f}$ denotes $\partial f/\partial r$ and $f'$ will denote $\partial f/\partial \phi$.
Any solution of the following first order equations is also a solution of the
second order field equations \cite{Skenderis:1999mm, DeWolfe:1999cp,Skenderis:2006jq}:
\be
\dot{A} = - \frac{1}{2 (d-1)} W; \qquad \dot{\phi} = W', \label{1st}
\ee
where the potential $V$ is written in the form:
\be
V = \frac{1}{2} \left [ (W')^2 - \frac{d}{2 (d-1)} W^2 \right ] \label{weq}
\ee
where $W$ is a superpotential. Note that $W$ does not necessarily need to coincide with the superpotential
${\cal W}$ of the supergravity theory into which the action (\ref{g-sc}) is embedded. Indeed the cases of
interest will be precisely those where the domain wall breaks the supersymmetry and admits only
fake supersymmetry \cite{Freedman:2003ax}. For the solution to correspond to
a vacuum expectation value (vev) the superpotential should have the
expansion around the AdS critical point $\phi=0$ of the form \cite{Papadimitriou:2004rz}
\be \label{vevform}
W=2 (d-1) +\frac{1}{2} \Delta \phi^2 + \cdots,
\ee
where $\Delta$ is the dimension of the dual operator that acquires
a vev\footnote{The other possible behavior is
$W = 2 (d-1) +1/2 (d-\Delta) \phi^2 + \cdots$. The corresponding domain wall
solution describes a deformation of the CFT by the dual operator.}.

The simplest example of a fake supersymmetric domain wall is obtained from a constant potential,
$V = V_0 = - d (d-1)$. The general solution for $W$ is then,
\be
W = 2 (d-1)  \cosh \left ( \sqrt{\frac{d}{2 (d-1)}} \phi \right ),
\ee
where we fixed the integration constant such that the AdS critical point is
at $\phi=0$. Near $\phi=0$ the superpotential has an expansion of
the form (\ref{vevform}) with $\Delta=d$ (which is indeed the dimension of the
dual operator since $\phi$ is massless).
Using this in the first order equations (\ref{1st}) results in
\bea
\phi &=& \sqrt{\frac{2 (d-1) }{d}} \ln \left [ \frac{1 - e^{-d (r - r^{\ast})}}{1 + e^{-d
(r - r^{\ast})}} \right ]; \label{back} \\
A &=& r + \frac{1}{d} \ln ( 1 - e^{-2d (r - r^{\ast})}), \nn
\eea
where $r^{\ast}$ is an integration constant. This solution is manifestly asymptotically $AdS_{d+1}$ as
$r \rightarrow \infty$ but there is a curvature singularity at $r = r^{\ast}$.

The dilaton domain wall solution for $d=4$ can be naturally embedded into
type IIB supergravity (where it was originally found by \cite{Kehagias:1999tr,Gubser:1999pk}) ; only the dilaton $\Phi$, the Einstein frame metric $g_{MN}$ and the five form $F$
are non-zero, with
\bea
ds^2 &=& L^2 \left [ e^{2 A (r)} \eta_{\mu \nu} dx^{\mu} dx^{\nu} + dr^2 \right ] + L^2 ds^2_{SE}; \\
F_5 &=& \frac{N \sqrt{\pi}}{2 V} (\eta_{SE} + \ast\eta_{SE}), \nn
\eea
with $\Phi = \phi(r)$. Here $ds^2_{SE}$ represents the metric on a Sasaki-Einstein manifold $X_5$,
with $\eta_{SE}$ its volume form and $V$ its volume. The components of the Einstein equations
along the compact directions are satisfied provided that $L^4 = \kappa_{10} N \sqrt{\pi}/2 V$.
The five-dimensional Newton constant defined in (\ref{g-sc}) is given by $2 \kappa_{5}^2 = 8 V/\pi N^2$.
This background is a solution for all Sasaki-Einstein
but breaks the supersymmetry since the dilatino transformation is always non-zero:
\be
\delta \chi = \frac{i}{2} \g^{M} \pa_{M} \Phi \ep^{\ast} + \cdots,
\ee
with ellipses denoting terms which vanish in this background. One needs to establish that this geometry
is dual to a symmetry breaking state (rather than a deformation); this can be done by computing the holographic
one point functions for the dual operators. Since the domain wall admits fake supersymmetry, one can show that it is perturbatively stable; indeed these
backgrounds were used to illustrate the implications of fake supersymmetry in \cite{Freedman:2003ax}.

The expectation values of the dual operators
can be extracted from the asymptotics of (\ref{back}) \cite{de Haro:2000xn} and are given by
\be \label{vevO}
\langle {\cal O} \rangle \sim
{\cal M}^d, \qquad \langle T_{ij} \rangle =0,
\ee
where ${\cal M} = e^{r^{\ast}}$, so ${\cal M}$ is the symmetry breaking scale.
Note however that the energy of the system is zero even though supersymmetry is broken.
This reflects the fact that supersymmetry cannot be broken spontaneously in superconformal
theories since the vanishing of the trace of the stress energy tensor is incompatible with having
non-zero energy. Furthermore, ${\cal M}$ is not fixed dynamically -- it is an integration
constant -- and its value can be changed without cost of energy.
This marginal direction is associated with the Goldstone
mode resulting from the spontaneously broken conformal symmetry. This
is also a manifestation of the general fact that one does not expect a
metastable state in a conformal field theory.

To obtain a more
realistic example, one could do the following: use a two scalar
potential, with one scalar corresponding to a supersymmetric relevant
operator in the dual CFT. Then construct a non-susy domain wall
solution in which this scalar has boundary conditions corresponding to
a supersymmetric deformation of the dual theory while the other scalar
has the slower falloff corresponding to a vev.
Such a domain wall could indeed correspond to a supersymmetry
breaking state in the dual supersymmetric non-conformal theory;
we postpone exploration of more
realistic examples of this type to future work.

Keeping in mind that the dilaton domain wall does not fully satisfy our requirements, let us nonetheless use it as a toy model to discuss how one would address the third required property of the holographic dual. The ${\cal N} = 2$ SCFT dual to an $AdS_5 \times X^5$ compactification with $X^5$ a Sasaki-Einstein admits a $U(1)_R$ symmetry associated with the Sasaki-Einstein Reeb vector as well as flavor symmetries associated with additional isometries of the $X^5$.
In the case of $S^5$ the dual theory is of course ${\cal N} = 4$ SYM with global symmetry $SO(6)$; for simplicity we will now
focus on this case, although the generalization to other Sasaki-Einsteins is straightforward.
Focussing on a single $U(1)$
flavor symmetry, the supergravity fields dual to the components of the current superfield (\ref{comp}) satisfy
linearized equations of motion:
\bea
J: & \qquad & \Box s = - 4 s; \\
j_{\alpha}: & \qquad & \gamma^{m} D_{m} \psi = \half \psi; \nn \\
j_{\mu}: & \qquad & D_{m} F^{mn} = 0. \nn
\eea
In the conformal vacuum the scalar $s$ is dual to the $\Delta = 2$  operator $J$; $\psi$ is dual to
the $\Delta = 5/2$ currents $j_{\alpha}$ whilst the gauge field $A_{m}$ with field strength $F_{mn}$ is dual to the
$\Delta = 3$ conserved current $j_{\mu}$. Whilst in general the equations of motion for fluctuations around the domain wall
background are complicated coupled equations, one can demonstrate that the modes dual to the components of the currrent superfield
actually satisfy free massive decoupled equations, of the same form as in the $AdS$ background. The
reason is that the active scalar in the background lies in a hypermultiplet of the ${\cal N} = 2$ gauged
supergravity whilst the modes dual to the currents lie in a vector multiplet. At the linearized level this suffices to show that
the scalar profile does not contribute to the field equations for the gauge field and its partners.

We now need to compute the current two point functions required for the generalized gauge mediation scenario via standard holographic
techniques. The logic proceeds as follows: one computes the renormalized holographic
one point function in the presence of a linearized source, substitutes the regular solution of the linearized field equation into
the one point function and differentiates with respect to the source to obtain the two point function, see e.g. \cite{Skenderis:2002wp}.
For example, for the scalar operator
$J$ the renormalized one point function is given in terms of the asymptotics of $s$ as $r \rightarrow \infty$
\bea
\< J \> &=& - {\cal N} \td{s}_{(0)}; \\
s(r,x) &=& e^{-2 r} r (s_{(0)} (x) + \cdots )  + e^{- 2 r} ( \td{s}_{(0)} (x) +  \cdots), \nn
\eea
where ellipses denote subleading terms as $r \rightarrow \infty$. The one point function
follows from differentiating with respect to the source
$s_{(0)} (x)$ the renormalized onshell action \cite{Bianchi:2001de,Bianchi:2001kw}
\bea
I &=&  {\cal N} \int d^5 x \sqrt{-g} \left ( \half (\partial s)^2 - 2 s^2 \right) + I_{ct}, \\
I_{ct} &=& {\cal N} \int_{r = | \ln \ep |}  d^4 x \sqrt{-h} \left ( 1 + \frac{1}{2 \ln \ep} \right ) s^2, \nn
\eea
where the counterterm action $I_{ct}$ at the regulated boundary $r =  | \ln \ep |$ renders
the action finite. The normalization factor ${\cal N}$ will be fixed below so that in the UV the
operator has unit normalization.
The field equation for $s (r,k)$ in momentum space is
\be
\left (\pa_y^2 + \coth(y) \pa_y -  \frac{q^2}{\sqrt{\sinh(y)}}  + \frac{1}{4} \right ) s(r,k) = 0,
\ee
where $y = 4 (r - r^{\ast})$ and $k^2/{\cal M}^2 = 16 \sqrt{2} q^2$. The scaling
behaviors of the terms in the equation imply it is not analytically solvable in terms of hypergeometric functions, so
to obtain the solution for general momentum numerical solution is necessary. One can however extract the zero and high momentum limits analytically, resulting in:
\be
\td{C}_0(0)= \frac{{\cal N}}{2} \ln 2, \qquad
\td{C}_0(k)_{k \rightarrow \infty} = - {\cal N} (\gamma + \ln (k/2)),
\ee
with $\gamma$ the Euler constant. Recalling that the regulated Fourier transform of $(1/|x|^4)$ is
$-\pi^2 \ln(k^2/\mu^2)$ one notes that the operator has unit normalization in the UV if ${\cal N} = 2 \pi^2$.

One can similarly compute the remaining coefficients in (\ref{con}), with the relevant differential equations being numerically solvable for general momentum whilst analytic analysis is possible at zero and high momentum:
\bea
\td{B}_{1/2}(0) &=& 0, \qquad \td{B}_{1/2}(k)_{k \rightarrow \infty} = 0; \\
\td{C}_{1/2}(0) &=& -\frac{{\cal N}}{\sqrt{2}} , \qquad \td{C}_{1/2}(k)_{k \rightarrow \infty} = \td{C}_0(k)_{k \rightarrow \infty}; \nn \\
\td{C}_1 (0) &=&  \frac{{\cal N}}{8} (\ln 2 -3), \qquad \td{C}_{1}(k)_{k \rightarrow \infty} = \td{C}_0(k)_{k \rightarrow \infty}, \nn
\eea
where we have identified the mass scale ${\cal M}$ in (\ref{vevO})
with the scale $M$ of (\ref{con}). These coefficients have the expected UV behavior.
Given the flaws of the toy model however the specific values of these coefficients are not to be taken seriously; in particular one does not find a non-zero value for
$\td{B}_{1/2}(0)$ when one imposes regularity conditions\footnote{{\bf Note added:} This point is further discussed in \cite{Bertolini}.}. However, one could follow the same approach to computing correlators in more realistic examples.
Note that integrating out the hidden sector also induces wavefunction renormalization
in the visible sector and produces a change in the beta-function of the gauge fields, $\delta b$, above the mass scale ${\cal M}$ as \cite{Meade:2008wd}.
If one wishes the visible sector to remain perturbative up to a GUT scale, the contribution to the beta function from the hidden sector is bounded. Reasonable choices of the GUT and susy breaking scales will constrain $| \Delta b |$ and correspondingly the value of $N$.

\section{Conclusions} \label{sec:disc}

In this work we have discussed a general approach to obtain holographic duals to metastable supersymmetry breaking states. One application of such holographic duals would be as realizations of hidden sectors in gauge mediated supersymmetric breaking and the geometries can be used to compute the two point functions required by the general gauge mediation scenario. We illustrated our approach with a toy example, the dilaton domain wall, and it would be interesting to explore the landscape of more realistic examples in future work.

\section*{Acknowledgments}

Preliminary versions of this work were discussed in talks in 2009, \cite{Taylor}. As this work was finalized we became aware of related work by R.~Argurio and collaborators \cite{Bertolini}; we thank these authors for bringing their forthcoming work to our attention. We would like to thank Elias Kiritsis, Costas Bachas and Adam Schwimmer for useful discussions.
This work is part of the research program of the `Stichting voor Fundamenteel Onderzoek der Materie (FOM)', which is financially
supported by the `Nederlandse Organisatie voor Wetenschappelijk
Onderzoek (NWO)'. KS also acknowledges support from NWO via a
Vici grant.

\end{document}